\documentclass[10pt, conference]{IEEEtran}
\IEEEoverridecommandlockouts
\usepackage{cite}
\usepackage{amsmath,amssymb,amsfonts,pifont}
\usepackage{algorithmic,multirow}
\usepackage{graphicx}
\usepackage{textcomp,soul}
\usepackage{xcolor,physics}
\usepackage[hyphens]{url}
\definecolor{myred}{RGB}{255,175,175}  

\def\BibTeX{{\rm B\kern-.05em{\sc i\kern-.025em b}\kern-.08em
    T\kern-.1667em\lower.7ex\hbox{E}\kern-.125emX}}
\begin{document}

\title{QuaLITi: \underline{Qua}ntum Machine \underline{L}earning Hardware Selection for \underline{I}nferencing with Top-\underline{Ti}er Performance
}

\author{\IEEEauthorblockN{Koustubh Phalak}
\IEEEauthorblockA{\textit{CSE Department} \\
\textit{Pennsylvania State University}\\
State College, PA \\
krp5448@psu.edu}
\and
\IEEEauthorblockN{Swaroop Ghosh}
\IEEEauthorblockA{\textit{School of EECS} \\
\textit{Pennsylvania State University}\\
State College, PA \\
szg212@psu.edu}}

\maketitle

\begin{abstract}
Quantum Machine Learning (QML) is an accelerating field of study that leverages the principles of quantum computing to enhance and innovate within machine learning methodologies. 
However, Noisy Intermediate-Scale Quantum (NISQ) computers suffer from noise that corrupts the quantum states of the qubits and affects the training and inferencing accuracy. Furthermore, quantum computers have long access queues. A single execution with a pre-defined number of shots can take hours just to reach the top of the wait queue, which is especially disadvantageous to Quantum Machine Learning (QML) algorithms that are iterative in nature. 
Many vendors provide access to a suite of quantum hardware with varied qubit technologies, number of qubits, coupling architectures, and noise characteristics.
However, present QML algorithms do not use them for the training procedure and often rely on local noiseless/noisy simulators due to cost and training timing overhead on real hardware. Additionally, inferencing is generally performed on reduced datasets with fewer datapoints. Taking these constraints into account, we perform a study to maximize the inferencing performance of QML workloads based on the choice of hardware selection. Specifically, we perform a detailed analysis of quantum classifiers (both training and inference through the lens of hardware queue wait times) on Iris and reduced Digits datasets under noise and varied conditions such as different hardware and coupling maps. We show that using multiple readily available hardware for training rather than relying on a single hardware, especially if it has a long queue depth of pending jobs, can lead to a performance impact of only 3-4\% while providing up to 45X reduction in training wait time. 
\end{abstract}


\section{Introduction}
In recent years, the field of quantum computing has witnessed significant growth, propelled by its potential to solve complex problems far beyond the reach of classical computing paradigms \cite{galindo2002information}. This emerging technology, characterized by its principles of superposition, entanglement, and quantum interference, offers unprecedented computational advantages, promising revolutionary breakthroughs \cite{arute2019quantum,kim2023evidence} in various disciplines, including cryptography \cite{mavroeidis2018impact}, finance \cite{orus2019quantum, herman2023quantum}, chemistry and material science \cite{bauer2020quantum}, and healthcare \cite{gupta2023quantum}. One of the most promising applications of quantum computing lies in the domain of machine learning, where the computational advantages of quantum algorithms can be leveraged to enhance the efficiency and capability of traditional machine learning algorithms \cite{biamonte2017quantum}. The synthesis of quantum computing and machine learning has given rise to a new interdisciplinary field known as Quantum Machine Learning (QML), which seeks to harness quantum computational advantages to improve machine learning tasks. Examples of quantum machine learning algorithms, such as Quantum Neural Networks (QNN) \cite{garg2020advances}, Variational Quantum Eigensolver (VQE) \cite{tilly2022variational}, Variational Quantum Classifier (VQC) \cite{cerezo2021variational}, and Quantum Support Vector Machine (QSVM) \cite{willsch2020support}, illustrate the potential of quantum computing to provide solutions to otherwise intractable learning problems.



\begin{figure}[t]
    \centering
    \includegraphics[width=0.8\linewidth]{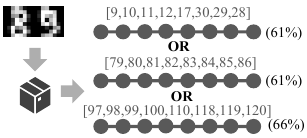}
    \caption{Training reduced Digits dataset (classes 8,9) on randomly allocated configurations gives poor inferencing results. The training is done on 127 qubit hardware where we observe a maximum inferencing performance up to $66\%$, suggesting that there is room for improvement with regards to the choice of qubit configuration and even hardware.}
    \label{fig:bad_config_training}
    \vspace{-5mm}
\end{figure}

\begin{figure*}[t]
    \centering
    \includegraphics[width=0.89\linewidth]{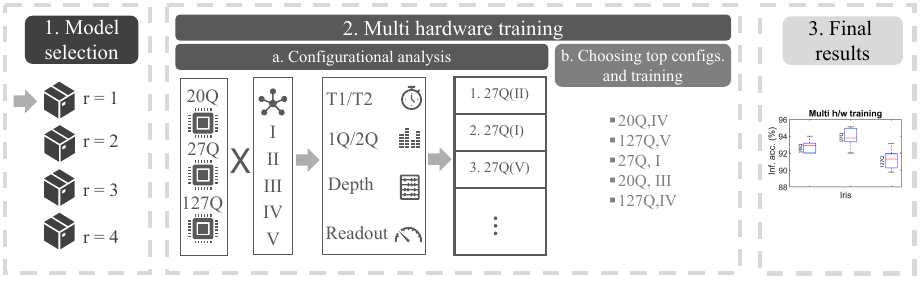}
    \caption{Main methodology for obtaining best inferencing results. We select 
    \raisebox{.5pt}{\textcircled{\raisebox{-.9pt} {\textbf{1}}}} suitable model that has the best entangling arrangement.  \raisebox{.5pt}{\textcircled{\raisebox{-.9pt} {\textbf{2}}}} The multi hardware training is performed as follows: \raisebox{.5pt}{\textcircled{\raisebox{-.9pt} {\textbf{a}}}} we perform configurational analysis based on various properties such as coherence times, error rates and circuit depth and give a cumulative score to each configuration. Then all the configurations are sorted by score. \raisebox{.5pt}{\textcircled{\raisebox{-.9pt} {\textbf{b}}}} The top scoring configurations from each hardware are selected for multi-hardware training. \raisebox{.5pt}{\textcircled{\raisebox{-.9pt} {\textbf{3}}}} 
    Inferencing is performed using the multi-hardware trained model and top performing values are noted. Note that I-V denote coupling maps as shown in Fig. \ref{fig:all_hw_coupling_map}(d).}
    \label{fig:methodology}
\end{figure*}

However, the practical realization of QML's potential is currently hindered by the limitations of the Noisy Intermediate-Scale Quantum (NISQ) technology era \cite{lau2022nisq} such as gate errors, decoherence, and crosstalk errors \cite{preskill2018quantum} and adhering to the hardware constraints e.g., coupling map that lead to performance degradation. 
Suppose, we want to train a QNN on a particular quantum hardware for performing binary classification. One iteration/epoch of training will constitute classifying all the data points by predicting their classes and using the predictions to compute the gradient and update the QNN parameters. Considering classifying even a single data point, the QNN will (i) undergo a transpilation process to make the circuit hardware compliant. This will increase the depth and gate count of the circuit, making it more susceptible to noise and, (ii) make multiple executions (also referred to as shots) to compute accurate expectation values for each datapoint. Repeating this process for all data points and multiple epochs can be challenging, given the NISQ constraints such as hardware wait times and noise levels. 
These constraints underscore the importance of hardware selection for QML workloads without significant degradation in training and inference accuracy. 

Previously, efforts have been made to mitigate the effects of some of the above constraints that include employing circuit concurrency in the QML training pipeline \cite{resch2021accelerating} (mitigating noise and reducing wait time), 
directly converting quantum circuits into native pulse schedules without the need for transpilation \cite{liang2022variational} (mitigating noise) and a similar work \cite{liang2024napa} (mitigating noise) using state preparation circuits catered to work robustly under noise \cite{wang2023robuststate} (mitigating noise), performing noise-aware training \cite{wang2022quantumnat} (mitigating noise) and training the unitary operator representation of a quantum circuit instead of the quantum circuit itself \cite{mate2022beyond} (mitigating noise).

\textbf{Motivation:} Suppose a user wants to train their QNN model on real quantum hardware, they accordingly define their QNN network consisting of an appropriate state preparation circuit to load classical data in quantum Hilbert space, the Parametric Quantum Circuit (PQC) consisting of trainable quantum layers and measurement operations for classical gradient computation and optimization of PQC parameters. However, suppose the user does not take the hardware constraints into account such as the coupling map and noise, and arbitrarily train their QNN without explicitly specifying a reasonably good configuration. In that case, there are chances that the QNN will be mapped to a poor configuration of qubits that can have high error rates, low coherence times, and high transpilation depth. For example, if a user wants to train their 8 qubit QNN using the Digits dataset (binary classification, classes 8 and 9) on 127 qubit hardware (qubits 0 to 126), with no configuration specified. There are chances that the QNN can be mapped to the following set of 8 qubits: (i) [9,10,11,12,17,30,29,28], (ii) [79,80,81,82,83,84,85,86] or (iii) [97,98,99,100,110,118,119,120]. These configurations have mean two qubit error rates up to as high as 0.31, and mean coherence times of all qubits as low as 70$\mu$s. These factors will corrupt the qubit states, leading to significant performance degradation. We can see that the inferencing performance reaches only a maximum of $66\%$ for all three configurations combined (Fig. \ref{fig:bad_config_training}). The situation can become even worse for multi-hardware training if the coupling maps are not chosen carefully. 

\textbf{Proposed approach:} In this work, we address the above concern by studying training of QNN on multiple hardware to maximize inferencing performance while cutting down wait time. We show the overall methodology in Fig. \ref{fig:methodology}. We choose the appropriate model, then perform configurational analysis to find out the best configurations and hardware and use them for multi-hardware training. Finally, we perform inferencing and obtain the results. Note that while in practice step 2 (multi-hardware training) should ideally run on real hardware, because of hardware access restrictions we perform step 2 using noisy simulations.
Although we use the Iris and reduced Digits datasets, the proposed methodology is generic and can be extended to larger datasets, such as a reduced Cifar-10 dataset as shown in the later part of this paper. 

In the rest of the paper, Section \ref{background_section} provides relevant background and related works. Section \ref{training_setup_section} explains the training setup followed by multi-hardware training setup and inferencing results in Section \ref{tr_inf_diff_hw_section}. Next, we perform additional analysis such as variation of inferencing performance, the effect of coupling map, real hardware queue depth analysis, and scalability to larger datasets in Section \ref{add_analysis_section}. Finally, we draw conclusions in Section \ref{conclusion_section}. All the code corresponding this work can be found in our GitHub repository\footnote{GitHub repository link: \url{https://github.com/KoustubhPhalak/QuaLITi-QML-Workload-Optimization}}. 

\section{Background and Related Works} \label{background_section}
\subsection{Quantum Computing}
Qubits are the fundamental units of a quantum computer, equivalent to bits for classical computers. A qubit stores information in a quantum state, which is represented using a 2x1 vector. It is mathematically denoted as
$\ket{\psi}=$
$\big[\begin{smallmatrix}
\alpha \\
\beta
\end{smallmatrix}\big]$, where  where $|\alpha|^2$ represents the probability of qubit being measured to 0 and $|\beta|^2$ represents the probability of qubit being measured to 1. There are two special states with $\alpha=1$, $\beta=0$ ( $\ket{0}=$
$\big[\begin{smallmatrix}
1 \\
0
\end{smallmatrix}\big]$) and $\alpha=0$, $\beta=1$ ( $\ket{1}=$
$\big[\begin{smallmatrix}
0 \\
1
\end{smallmatrix}\big]$). These are known as basis states and are quantum analogous to classical 0 and 1 bit values respectively. The quantum state of a qubit is changed with the help of quantum gates, which are unitary matrix operations. These gates work either on a single qubit (e.g., Hadamard gate, Pauli X/Y/Z gates, etc.) or on multiple qubits (CNOT gate, Pauli CY/CZ gate, SWAP gate, Toffoli gate, etc.). Combining qubits and quantum gates, we obtain quantum circuits that are ordered sequences of quantum gates placed on qubits. All the gate operations of qubits are eventually collapsed to classical bit values (either 0 or 1) \cite{von2018mathematical}. A special kind of quantum circuit, known as Parametric Quantum Circuit (PQC) consists of parametric rotation gates (such as U, Pauli RX/RY/RZ, etc.) that can be tuned classically using traditional optimization algorithms. These PQCs can be thought of as trainable ML models that form an integral part of QNNs in QML.

\subsection{Cloud-based NISQ Quantum Computing}
Modern NISQ quantum computers today are typically accessed via a cloud service from vendors such as IBM \cite{IBMQuantum}, Google \cite{GoogleQuantumComputer}, and Amazon \cite{AmazonBraket}. The general way to access a quantum computer is (1) the user would write the program containing the specific quantum circuit for their task, (2) the user would then send the program to the cloud service along with the target hardware to run the quantum circuit on and some extra metadata (such as number of shots, optimization level for transpilation, number of ancilla qubits, etc.), (3) the cloud service would allocate the user's program to the desired hardware, after which the hardware would run the program, generate the results and send them back to the cloud service, (4) finally, the cloud service sends the results from the quantum hardware back to the user. 

However, there is a major problem that arises from such an access model. Due to the limited availability of quantum computers, a single quantum computer has a long access queue that can have up to hours of wait time until the user's program reaches the top of the queue. 
This problem is further aggravated for QML algorithms, that require iterative runs for optimizing the rotation gate parameters for which
selecting a quantum hardware that has less wait time is an important aspect for consideration in the QML training pipeline.

\subsection{Noise in quantum hardware}
The performance of QNNs, like all quantum computing systems, is significantly influenced by various forms of noise, each impacting the accuracy and efficiency of the system in distinct ways. A common source of error within QNNs is decoherence, a phenomenon where qubits lose their quantum state due to unintended interactions with the external environment. This loss is essentially an energy dissipation from the qubits, leading to a degradation of the quantum coherence of the system and, consequently, its computational capabilities. Another error is crosstalk which occurs when there are unwanted interactions specifically between qubits that are quantum mechanically coupled. These interactions can alter the state of neighboring qubits in an uncontrolled manner, introducing errors into the computation process. 

The implementation of quantum gates also introduces potential sources of error. Quantum gates in QNNs are typically realized through the application of microwave pulses in systems utilizing superconducting qubits, or laser pulses in the case of trapped-ion qubits. Any inaccuracies or imperfections in these pulses can result in gate errors, where the intended quantum operation is applied incorrectly, leading to deviations from the expected computational outcome. Lastly, the process of measuring quantum states introduces another avenue for error, known as readout errors. Quantum measurement operations can vary widely depending on the physical implementation of the qubits. For instance, photonic qubits are often measured using photon detectors, trapped-ion qubits may be measured through the intensity of fluorescence, and superconducting qubits might be measured via resonator coupling. Each measurement technique has its own set of potential inaccuracies, whether due to imprecise measurements, equipment limitations, or inherent inaccuracies in the measurement apparatus itself. These readout errors can significantly affect the accuracy of the quantum computation, as they directly influence the interpretation of the quantum system's final state.

\subsection{Related Works}
Many efforts have been attempted to reduce the effect of quantum hardware limitations. For example, \cite{resch2021accelerating} was proposed to run concurrent executions of different training data within the same batch for the same QNN circuit on different available qubits. \cite{liang2022variational,liang2024napa} propose converting Variational Quantum Algorithms (VQA) into pulse schedules such that the pulses are native to the quantum hardware and have tunable parameters. By training these parameters, the authors train the original model. Works like \cite{wang2023robuststate,wang2022quantumnat} take the effect of noise into account and incorporate them into the VQA, such as creating noise-resilient state preparation circuit for better training, and injecting quantum noise during training to make it noise-aware and performing post-measurement processing, such as quantization and normalization of measurement outputs. Finally, \cite{mate2022beyond} propose training unitary operator representation of a QML ansatz ($2^N*2^N$ in size for $N$ qubits) as compared to training the ansatz itself. The authors use a gradient descent algorithm for optimization and also use partitioning of the unitary operator for further time complexity reduction. Out of all these works, only \cite{resch2021accelerating} takes hardware queue wait time into consideration in its study and achieves up to 20x speedup. The proposed approach is complementary in nature and could be augmented in conjunction with \cite{resch2021accelerating} for additional benefit.


\section{Training Setup} \label{training_setup_section}
\subsection{Hardware and configuration selection}
\begin{figure}[t]
    \centering
    \includegraphics[width=\linewidth]{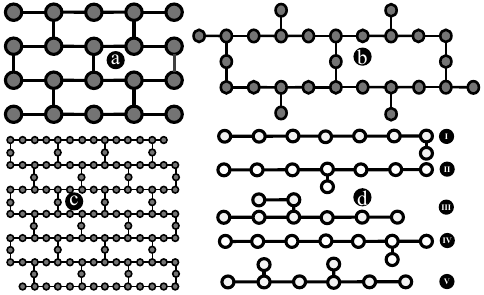}
    \caption{Coupling maps of (a) 20 qubit hardware, (b) 27 qubit hardware and (c) 127 qubit hardware; (d) five coupling configurations I-V used for noisy experiments.}
    \label{fig:all_hw_coupling_map}
    \vspace{-4mm}
\end{figure}

We first choose a set of quantum hardware and their corresponding configurations for training and inferencing QML models. Considering the latest release of Qiskit \textit{1.0}, we select \textit{Fake20QV1()} (20 qubits), \textit{Fake27QPulseV1()} (27 qubits) and \textit{Fake127QPulseV1()} (127 qubits) noisy simulators (containing real hardware calibrated data) from \textit{qiskit.providers.fake\_provider} library for conducting our experiments. The coupling map of each of these hardware is shown in Fig. \ref{fig:all_hw_coupling_map}. We select five topologically different 8 qubit coupling maps (labeled I-V) for each hardware as shown visually in Fig. \ref{fig:all_hw_coupling_map}(d) and with individual qubit information in Table \ref{tab:coupling_configs}. Considering the exponential simulation runtime increase with growing qubits \cite{kitaev2000parallelization, daley2022practical}, we pick small-scale datasets such as the Iris and UCI Digits dataset for the training process. Furthermore, we select all 3 classes in the Iris dataset for classification and 2 sets of 2 class datasets in the Digits dataset: 0,1 and 8,9 for performing classification\footnote{Henceforth, we refer to them as Digits01 and Digits89 respectively.}.  

\begin{table}[t]
\centering
\caption{8-qubit coupling configurations used in different hardware. Note that C.M. = Coupling Map}
\label{tab:coupling_configs}
\begin{tabular}{|l|l|l|l|}
\hline
\textbf{C.M.} &
  \textbf{27Q} &
  \textbf{127Q} &
  \textbf{20Q} \\ \hline
I &
  \begin{tabular}[c]{@{}l@{}}{[}4,7,10,12,\\ 15,18,20,23{]}\end{tabular} &
  \begin{tabular}[c]{@{}l@{}}{[}14,18,19,20,\\ 21,22,23,24{]}\end{tabular} &
  \begin{tabular}[c]{@{}l@{}}{[}0,1,6,5,\\ 10,11,15,16{]}\end{tabular} \\ \hline
II &
  \begin{tabular}[c]{@{}l@{}}{[}4,7,10,12,\\ 13,15,18,20{]}\end{tabular} &
  \begin{tabular}[c]{@{}l@{}}{[}19,20,21,22,\\ 23,24,25,15{]}\end{tabular} &
  \begin{tabular}[c]{@{}l@{}}{[}5,6,7,8,\\ 9,14,13,3{]}\end{tabular} \\ \hline
III &
  \begin{tabular}[c]{@{}l@{}}{[}7,10,12,15,\\ 18,20,13,14{]}\end{tabular} &
  \begin{tabular}[c]{@{}l@{}}{[}20,21,22,23,\\ 24,25,15,4{]}\end{tabular} &
  \begin{tabular}[c]{@{}l@{}}{[}6,7,8,9,\\ 14,13,3,2{]}\end{tabular} \\ \hline
IV &
  \begin{tabular}[c]{@{}l@{}}{[}10,12,15,18,\\ 20,23,24,13{]}\end{tabular} &
  \begin{tabular}[c]{@{}l@{}}{[}14,18,19,20,\\ 21,22,23,15{]}\end{tabular} &
  \begin{tabular}[c]{@{}l@{}}{[}0,1,2,3,\\ 8,9,14,6{]}\end{tabular} \\ \hline
V &
  \begin{tabular}[c]{@{}l@{}}{[}4,7,10,12,\\ 15,18,6,13{]}\end{tabular} &
  \begin{tabular}[c]{@{}l@{}}{[}19,20,21,22,\\ 23,24,15,33{]}\end{tabular} &
  \begin{tabular}[c]{@{}l@{}}{[}5,6,1,2,\\ 3,4,0,8{]}\end{tabular} \\ \hline
\end{tabular}
\vspace{-4mm}
\end{table}

\subsection{Model selection}


We use 8-qubit QNNs containing classical data loading circuits, Parametric Quantum Circuit (PQC), and measurement operations. 
We select the input data loading scheme based on the dimensions of the input. Considering we have 8 qubit-QNN, the Iris dataset ($4$ size vector) can fit with an $n$ features to $n$ qubits embedding, while the Digits dataset ($64$ size vector) can fit with a $2^n$ features to $n$ qubits embedding. So, we use angle embedding ($4$ features on $4$ qubits) for the Iris dataset and amplitude embedding ($64$ features on $6$ qubits) for the Digits dataset \cite{nielsen2001quantum}.
Strongly Entangling Layers (SEL) \cite{schuld2020circuit} are selected for the PQC part since they create strong entanglement due to the presence of many entangling gates such as CNOT gates, and also have many trainable parameters which overall improves trainability of the PQC \cite{cai2015entanglement}. Furthermore, we take the number of classes into account and accordingly set the number of layers for each case. Specifically for the Iris dataset (with 3 classes) we choose a model with 6 SEL and for the reduced Digits datasets (with 2 classes) we choose a model with 3 SEL. Finally, the measurement operations consist of expectation value measurement in the Pauli-Z basis. For all the runs, we perform training on 10 epochs, use Adam optimizer (learning rate $=10^{-3}$), and use a batch size of 16.

\begin{figure}[t]
    \centering
    \includegraphics[width=\linewidth]{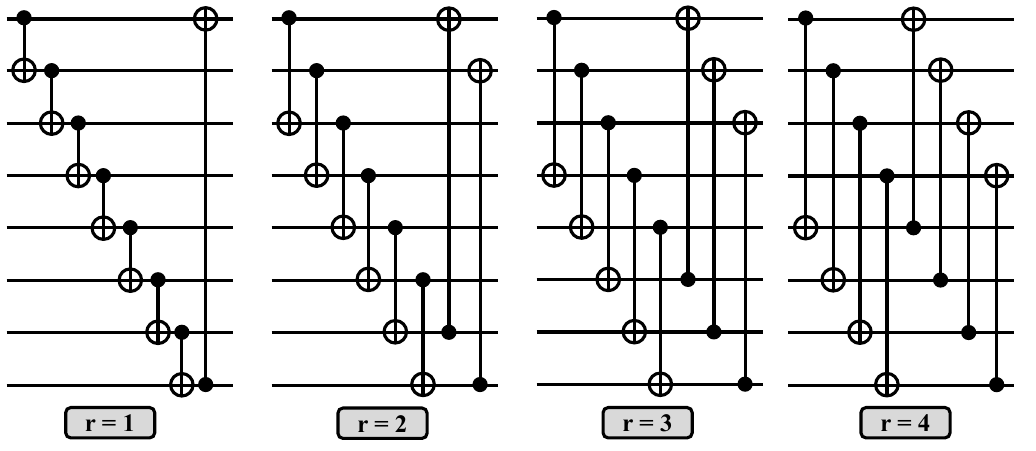}
    \caption{Different arrangements of CNOT gates in SEL.}
    \label{fig:cnot_arrangement}
    \vspace{-4mm}
\end{figure}

\begin{table}[t]
\centering
\caption{Inferencing performance of models with varying $r$.}
\label{tab:model_selection}
\begin{tabular}{|l|l|l|l|}
\hline
\textbf{Range (r)} & \textbf{Iris} & \textbf{Digits01} & \textbf{Digits89} \\ \hline
1                  & 93.78\%       & 89.35\%           & 84.57\%           \\ \hline
2                  & 86.00\%       & 84.72\%           & 83.73\%           \\ \hline
3                  & 88.44\%       & 90.74\%           & 80.37\%           \\ \hline
4                  & 74.22\%       & 61.38\%           & 75.23\%           \\ \hline
\end{tabular}
\vspace{-4mm}
\end{table}

\begin{table*}[t]
\centering
\caption{Analysis of various configurations
}
\label{tab:config_analysis}
\begin{tabular}{|l|l|l|l|l|l|l|}
\hline
\textbf{Config.} & \textbf{\begin{tabular}[c]{@{}l@{}}Decoherence score\\ normalized (A)\end{tabular}} & \textbf{\begin{tabular}[c]{@{}l@{}}Readout score\\ normalized (B)\end{tabular}} & \textbf{\begin{tabular}[c]{@{}l@{}}1Q error score\\ normalized (C)\end{tabular}} & \textbf{\begin{tabular}[c]{@{}l@{}}2Q error score\\ normalized (D)\end{tabular}} & \textbf{\begin{tabular}[c]{@{}l@{}}Layer depth score\\ normalized (E)\end{tabular}} & \textbf{\begin{tabular}[c]{@{}l@{}}Final score=0.2(A)+0.2\\ (B)+0.1(C)+0.3(D)+0.2(E)\end{tabular}} \\ \hline
27Q(II)          & 90.23                                                                   & 83.91                                                                   & 100.00                                                                             & 93.80                                                                      & 10.47                                                                   & \textit{75.06}                                                                                 \\ \hline
27Q(I)           & 76.91                                                                   & 85.59                                                                   & 65.91                                                                      & 100.00                                                                             & 19.47                                                                   & \textit{72.98}                                                                                 \\ \hline
27Q(V)           & 100.00                                                                          & 100.00                                                                           & 100.00                                                                             & 68.69                                                                      & 2.08                                                                   & \textit{71.02}                                                                                 \\ \hline
27Q(III)         & 89.06                                                                   & 89.01                                                                   & 65.91                                                                      & 86.23                                                                      & 1.02                                                                   & \textit{68.28}                                                                                 \\ \hline
27Q(IV)          & 47.97                                                                   & 75.94                                                                    & 65.91                                                                      & 76.13                                                                      & 13.26                                                                   & \textit{56.86}                                                                                 \\ \hline
20Q(II)          & 3.69                                                                   & 29.04                                                                   & 9.09                                                                      & 13.78                                                                      & 100.00                                                                           & \textit{31.59}                                                                                 \\ \hline
127Q(V)          & 5.55                                                                   & 64.38                                                                   & 11.36                                                                      & 15.28                                                                      & 51.51                                                                   & \textit{30.01}                                                                                 \\ \hline
20Q(IV)          & 0.57                                                                   & 32.71                                                                   & 13.46                                                                      & 8.30                                                                      & 73.83                                                                     & \textit{25.26}                                                                                 \\ \hline
127Q(II)         & 10.11                                                                   & 64.38                                                                   & 11.36                                                                      & 22.50                                                                      & 28.72                                                                   & \textit{28.53}                                                                                 \\ \hline
127Q(I)          & 7.58                                                                   & 56.56                                                                   & 11.36                                                                      & 29.71                                                                      & 14.73                                                                   & \textit{25.82}                                                                                 \\ \hline
127Q(IV)         & 6.68                                                                   & 51.70                                                                   & 14.77                                                                      & 19.58                                                                      & 24.80                                                                   & \textit{23.98}                                                                                 \\ \hline
20Q(III)         & 0.00                                                                            & 22.19                                                                   & 7.95                                                                      & 9.60                                                                      & 54.73                                                                   & \textit{19.06}                                                                                 \\ \hline
20Q(I)           & 12.33                                                                   & 0.00                                                                             & 0.00                                                                               & 14.77                                                                      & 73.83                                                                     & \textit{21.66}                                                                                 \\ \hline
20Q(V)           & 7.55                                                                   & 34.61                                                                   & 13.46                                                                      & 0.00                                                                               & 48.46                                                                   & \textit{19.47}                                                                                 \\ \hline
127Q(III)        & 9.55                                                                   & 18.99                                                                   & 10.51                                                                      & 20.15                                                                      & 0.00                                                                             & \textit{12.80}                                                                                 \\ \hline
\end{tabular}
\vspace{-5mm}
\end{table*}

Another aspect to consider in the model selection process is the ansatz used in SEL. The SELs have a range parameter $r$ that can dictate the target qubit for a given control qubit of the CNOT gate. Suppose the control qubit is on $i^{th}$ qubit (ranging $0$ to $n-1$), the total number of qubits in the QNN is $n$ and the range value is $r$, then the target qubit will be on qubit number $(i+r)$(mod $n$). As an example, we show for an 8 qubit SEL (entangling part in particular) how having different range values changes the arrangement of CNOT gates in Fig. \ref{fig:cnot_arrangement}. We select four range values $r=\{1,2,3,4\}$, define a model for each range value and train these models under noise using the chosen datasets to determine the model with most appropriate entanglement. For each case, the training is done on 27 qubit hardware with configuration I (linear coupling map). We note the inferencing results for this experiment in Table \ref{tab:model_selection}. Note that the tabulated values present mean inferencing performance for 10 inferencing runs to account for fluctuations due to noise\footnote{For all the runs, we use mean accuracy of 10 inferencing runs.}. We observe that for the Iris dataset, $r=1$ performs the best, and for Digits01 and Digits89 $r=3$ and $r=1$ perform the best, respectively. Therefore, we select these models for further analysis.

\section{Multi hardware training} \label{tr_inf_diff_hw_section}
\subsection{Configurational analysis}
We identify the potentially best-performing configurations by analyzing their various properties such as coherence times (both T1, and T2 times), single and two-qubit error rates, single-layer post-transpilation depth, and readout error. In general, the presence of CNOT gates due to the SWAP gate insertion procedure during transpilation makes the resulting QNN circuit sensitive to the two-qubit error rate. Next, the native gate set of a particular hardware plays an important factor in determining the overall QNN depth. 
The readout errors (preparing $\ket{0}$, measuring $1$ and preparing $\ket{1}$, measuring 0) can further degrade the performance. Finally, individual single qubit error rates can lead to the erroneous computation of quantum state, however, their effect can be considered relatively small compared to other factors. For the best results, (i) the \textit{two qubit error rate}, \textit{post-transpilation depth}, \textit{readout errors} and \textit{single qubit error rate} should be low (inversely proportional $\downarrow$), and (ii) the \textit{coherence times} should be high (directly proportional $\uparrow$). Furthermore, both the types of aforementioned readout errors should be low and both T1 and T2 times should be high. Taking this into account, we can combine the T1 and T2 times by taking their harmonic means and the readout errors using arithmetic mean. 
The harmonic mean is sensitive to lower values, so a pair of T1, and T2 values having even a single low value will have a lower harmonic mean, implying a lower overall coherence time for the configuration. Similarly, high readout error values will yield a higher arithmetic mean. Next, we create a score metric for every property in the range of [0,100]. We do this as follows: (i) For inversely proportional properties, we take the inverse/reciprocal of the property and normalize them in the range of [0,100]. The normalization is done by taking the minimum and maximum of the property value into account. Suppose, if $p$ is the list of all the scores of the property after taking inverse, then for a property score $p_i$ ($1\leq i\leq15$ since there are 5 configurations for 3 hardware), the normalized score will be $p_i'=\frac{p_i - min(p)}{max(p)-min(p)}*100$. (ii) For directly proportional values, we directly normalize the property score value in the range [0,100] using the aforementioned formula without taking the inverse. 
Finally, once the individual score for each property is computed, we combine these individual scores in a weighted fashion to obtain a final overall score for each configuration. Based on the criticality of the properties discussed earlier, we assign a weight of 0.3 to two qubit error rate, 0.2 to post-transpilation layer depth, coherence times, and readout error rates, and 0.1 to single-qubit error rate. These results are tabulated in Table \ref{tab:config_analysis} in decreasing order of final score from top to bottom. From the table, we observe that all the 27 qubit hardware configurations are top performing, owing to high coherence times and lower error rates.

\subsection{Multi hardware training procedure} \label{multi_hw_section}
From the analysis performed in Table \ref{tab:config_analysis}, it would be motivating to select the top-scoring configurations with the highest scores. However, as we can see the top 5 scoring configurations all belong to the 27 qubit hardware. From a training standpoint,
while the individual coupling map configurations might be different we still dedicate all 10 epochs of training only to 27 qubit hardware. If the 27 qubit hardware has a large queue of pending jobs, then the overall training time overhead will be equivalently compounded based on the number of epochs allotted to it for training. To address this challenge, we propose an alternative configuration selection strategy to just selecting the top-scoring configurations, where 
(i) We select one of the top-scoring configurations from each hardware at least once during training, and (ii) the next hardware selected for training should be different from the current hardware chosen.
We specifically use these two criteria since they allow the usage of top-performing hardware and avoid the case of choosing the busiest hardware back to back, potentially saving training wait time in the hardware queue. We satisfy these two conditions by selecting five configurations (2 epochs per configuration) such that the first three configurations are among the best-performing configurations from each hardware and the next two 
are chosen in a similar fashion but 
we also ensure that these configurations have the least queue wait times.
For example, the top three scored configurations (in order) for each hardware are (i) 20Q: II, IV, III (ii) 27Q: II, I, V (iii) 127Q: V, II, IV. From these, we can randomly select one configuration from each hardware, switch to a different hardware, and repeat this process 5 times. We employ randomness in the selection here as the scores of specified configurations for each hardware are relatively close so selecting one configuration over the other will not make much difference in the final inferencing performance. A set of configurations chosen in this fashion for the Iris dataset could be 20Q (IV), 127Q (V), 27Q (I), 20Q (III), and 127Q (IV).
The final training configurations for all the datasets selected using this procedure are shown in Table \ref{tab:multi_hw_all_hw_configs} in order of training from left to right. Note that we assume all the chosen configurations have the least wait time.

\begin{figure}[t]
    \centering
    \includegraphics[width=\linewidth]{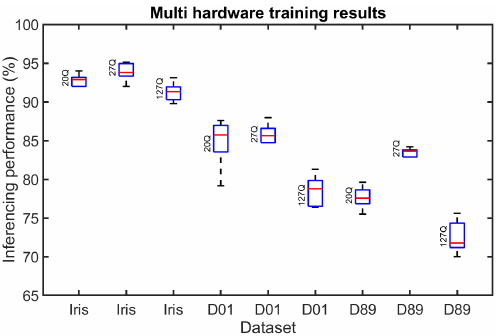}
    \caption{Multi hardware training results. D01 = Digits01, D89 = Digits89.}
    \label{fig:multi_hw_results}
    \vspace{-4mm}
\end{figure}

The results for this training procedure are shown in Fig. \ref{fig:multi_hw_results}. We observe that 27 qubit hardware is the best-performing hardware for inferencing, followed by 20 qubit hardware and finally 127 qubit hardware. 
For 20, 27 and 127 qubit hardware respectively, we note mean inferencing accuracy of (i) $92.76\%$, $93.91\%$ and $91.24\%$ for Iris dataset, (ii) $84.85\%$, $85.83\%$ and $78.47\%$ for Digits01 dataset and (iii) $77.65\%$, $83.45\%$ and $72.57\%$ for Digits89 dataset.


\begin{table}[t]
\centering
\caption{Configurations selected for multi-hardware training procedure}
\label{tab:multi_hw_all_hw_configs}
\begin{tabular}{|l|lllll|}
\hline
\textbf{Dataset} & \multicolumn{5}{c|}{\textbf{Configurations (in order from   left to right)}} \\ \hline
Iris     & \multicolumn{1}{l|}{20Q, IV}    & \multicolumn{1}{l|}{127Q, V} & \multicolumn{1}{l|}{27Q, I} & \multicolumn{1}{l|}{20Q, III}   & 127Q, IV \\ \hline
Digits01 & \multicolumn{1}{l|}{127Q, II} & \multicolumn{1}{l|}{27Q, I}  & \multicolumn{1}{l|}{20Q, III}  & \multicolumn{1}{l|}{27Q, V} & 20Q, IV   \\ \hline
Digits89 & \multicolumn{1}{l|}{27Q, II}   & \multicolumn{1}{l|}{20Q, II}  & \multicolumn{1}{l|}{127Q, V}  & \multicolumn{1}{l|}{27Q, I}  & 20Q, IV   \\ \hline
\end{tabular}
\vspace{-3mm}
\end{table}

\begin{table}[t]
\centering
\caption{Mean hardware characteristics of different hardware for all coupling configurations combined. Note that A=\{id,u1,u2,u3,cx\} and B=\{id,rz,sx,x,cx,reset\}.}
\label{tab:hw_characteristics}
\begin{tabular}{|l|l|l|l|}
\hline
\textbf{Property} & \multicolumn{1}{c|}{\textbf{20Q}} & \textbf{27Q} & \textbf{127Q} \\ \hline
2q error rate     & 0.0172                           & 0.0085       & 0.0147        \\ \hline
Basis gate set    & A                                 & B            & B             \\ \hline
\end{tabular}
\vspace{-5mm}
\end{table}


\section{Additional Analysis} \label{add_analysis_section}
\subsection{Performance variation with hardware and dataset}
\textbf{Variation with hardware:} From Fig. \ref{fig:multi_hw_results}, we observe inferencing performance variation as we switch to different hardware. In particular, we note that 27 qubit hardware performs the best, followed by 20 qubit hardware and finally the worst-performing 127 qubit hardware. This trend can be explained by examining internal hardware characteristics.

We show the mean hardware characteristics for all coupling configurations combined such as mean two qubit error rates in Table \ref{tab:hw_characteristics} and the native basis gate set for every hardware.
We observe that 27 qubit hardware has the best two-qubit error rate (0.0085) (which matches with the high two-qubit error scores of all 27 qubit hardware in Table \ref{tab:config_analysis}). 
The 20-qubit and 127-qubit hardware have two-qubit error rates that are a magnitude of order higher (0.0172 and 0.0147 respectively) than the 27-qubit hardware (which explains why it shows the best performance). Furthermore, the 20 qubit hardware has a different basis gate set compared to the other two, which leads to lower post-transpilation depth. For example, an amplitude embedding circuit along with a single SEL with r=1 post-transpilation has an average depth of 608 on 127 qubit hardware while having 321 on 20 qubit hardware. Therefore, even if the two-qubit error rate is higher, the lower post-transpilation depth compensates for the error rate and gives higher performance for 20-qubit hardware, as compared to the 127-qubit hardware. However, if 27 qubit hardware has a large queue, then we may lose performance by selecting other hardware in the process of optimizing wait time. This is a trade-off between the wait time and inferencing performance that should be made while choosing the desired configurations.

\textbf{Variation with dataset:} As mentioned earlier, we utilize angle embedding to embed a vector of size 4 onto 4 qubits using parametric rotation angles for the Iris dataset, and amplitude embedding to embed 64 size feature vectors onto 6 qubits. Under ideal conditions, both the state preparation circuits perform well. However, when coupling constraints of hardware are taken into account, an amplitude embedding circuit requires depth exponential in the number of qubits used \cite{weigold2020data}. This contributes significantly to the overall post-transpilation circuit depth, leading to degradation in performance. This can be seen when we compare the performances of the Iris and Digits dataset in Fig. \ref{fig:multi_hw_results}. We note that Iris dataset always shows greater than $90\%$ inferencing performance for any configuration, and Digits dataset shows inferencing performance below $90\%$, sometimes even below $80\%$.

We also observe significant performance differences within the two Digits datasets. Structurally, both datasets have images of the same dimensions (8x8 images, or 64x1 size vector for the case of training). This means that both datasets will have the same amplitude embedding depth post-transpilation. Numerically, we observe a post-transpilation depth for the amplitude embedding circuit of roughly 490 for both datasets. The remaining difference then would be the digits present in the images themselves, specifically the structure of the digits. For the case of Digits01, 0 has a round `o' shape while 1 has line `$|$' shape, while for Digits89 digit 8 has two `o' shapes while 9 has one `o' shape attached to a `$|$' shape. Under noiseless conditions, the selected model can classify both datasets with greater than $90\%$ accuracy. However, under noise, the same model would probably find it easier to distinguish between images containing two distinct shapes (`o' and `$|$') like 0 and 1, as compared to images having common shapes (`o' shape) like 8 and 9. This can potentially explain the lower performance in the Digits89 dataset as compared to the Digits01 dataset. 

\begin{figure}[t]
    \centering
    \includegraphics[width=\linewidth]{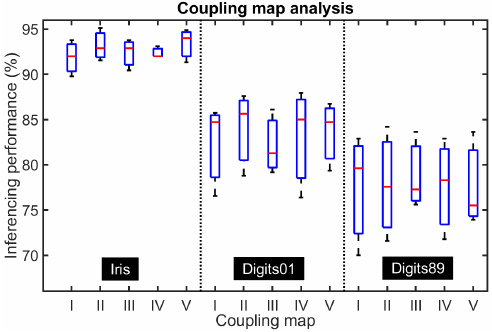}
    \caption{Variation in inferencing performance with different coupling maps for multi-hardware training setup.}
    \label{fig:cm_variation}
    \vspace{-3mm}
\end{figure}

\begin{table*}[t]
\centering
\caption{Training and inferencing wait times on real hardware}
\label{tab:train_inf_wait_time_table}
\begin{tabular}{|l|l|l|l|l|l|l|l|}
\hline
\multirow{2}{*}{\textbf{H/W}} &
  \multirow{2}{*}{\textbf{\begin{tabular}[c]{@{}l@{}}Queue wait\\ time (A)\end{tabular}}} &
  \multirow{2}{*}{\textbf{\begin{tabular}[c]{@{}l@{}}Queue\\ depth (B)\end{tabular}}} &
  \multirow{2}{*}{\textbf{\begin{tabular}[c]{@{}l@{}}Avg wait time\\ C = (A $\divisionsymbol$ B)\end{tabular}}} &
  \multirow{2}{*}{\textbf{\begin{tabular}[c]{@{}l@{}}\# Train data\\ (D, Iris/Digits)\end{tabular}}} &
  \multirow{2}{*}{\textbf{\begin{tabular}[c]{@{}l@{}}Train wait time\\ (C*D*10$\divisionsymbol$60)\end{tabular}}} &
  \multirow{2}{*}{\textbf{\begin{tabular}[c]{@{}l@{}}\# Inf. data\\ (E, Iris/Digits)\end{tabular}}} &
  \multirow{2}{*}{\textbf{\begin{tabular}[c]{@{}l@{}}Inf. wait \\ time (C*E)\end{tabular}}} \\
         &        &      &       &         &              &        &            \\ \hline
Brisbane & 5h 40m & 1447 & 14.1s & 105/252 & 247/592 min  & 45/108 & 634/1522s  \\ \hline
Kyoto    & 2h 15m & 183  & 44.3s & 105/252 & 775/1860 min & 45/108 & 1993/4784s \\ \hline
Osaka    & 1s     & 1    & 1s    & 105/252 & 17.5/42 min  & 45/108 & 45/108s    \\ \hline
\end{tabular}
\vspace{-4mm}
\end{table*}

\subsection{Effect of changing coupling map}
We visualize previous results from the viewpoint of coupling maps. 
We restructure multi-hardware training results from Fig. \ref{fig:multi_hw_results} in Fig. \ref{fig:cm_variation} in the form of boxplots for each coupling map and every dataset.
An indicator of a good coupling map configuration is the lower fluctuation it shows across different hardware.  
Across all the three datasets combined for coupling maps I to V respectively, we observe mean fluctuations of $3.74\%$, $3.46\%$, $2.59\%$, $3.33\%$, and $2.96\%$. From these values, we can conclude that coupling maps III, and V are the most resilient for inferencing multi-hardware trained models.

\begin{figure}[t]
    \centering
    \includegraphics[width=\linewidth]{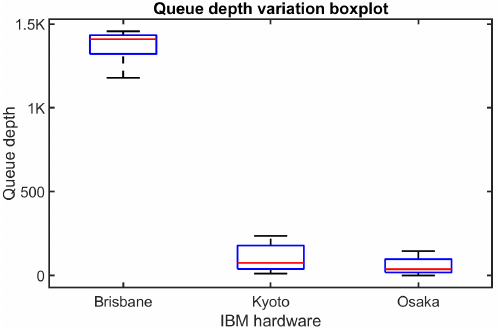}
    \caption{Queue depth boxplots for different IBM quantum hardware.}
    \label{fig:queue_plots}
    \vspace{-4.5mm}
\end{figure}



\subsection{Hardware queue depth analysis}
We observe queue depth variation of real hardware (IBM Brisbane, IBM Kyoto and IBM Osaka) with time. We show boxplot for queue depth (recorded from 04/10/2024 16:30 to 04/12/2024 19:30) in Fig. \ref{fig:queue_plots}. We note that IBM Brisbane is the busiest out of all three, followed by IBM Kyoto and IBM Osaka. From the boxplots, we note the overall fluctuation in queue depth (i.e. standard deviation) as 88 jobs for IBM Brisbane, 74 jobs for IBM Kyoto, and 43 jobs for IBM Osaka. To get an idea of how much wait time these queue depths translate to, we choose a dummy 8-qubit circuit (consisting of only a single RZ gate on every qubit) and send it for execution on each circuit. The queue wait times for this circuit is tabulated in Table \ref{tab:train_inf_wait_time_table}. We observe that for the aforementioned time period, Kyoto has the largest wait time per circuit (44.3s), followed by Brisbane (14.1s) and Osaka (1s). Using this data, we estimate the single inference run time. For our datasets, we select a 70:30 split for training and testing, which would mean (for inferencing) 45 data points for the Iris dataset and 108 data points for both the Digits01 and Digits89 datasets. Assuming each datapoint run would wait for the computed wait time, we obtain an overall inferencing wait time of 634, 1993, and 45 seconds on IBM Brisbane, IBM Kyoto, and IBM Osaka respectively for the Iris dataset and 1522, 4784, and 108 seconds respectively for both the Digits datasets for a single inferencing run. We can also extrapolate the average wait time to estimate training wait time. For example, consider the Iris dataset for training on the IBM Osaka machine. The training set will have 105 images, so for an average wait time of 1s and 10 epochs of training, the total training wait time will be roughly 17 minutes. We tabulate all the training wait times for all datasets on all hardware in Table \ref{tab:train_inf_wait_time_table}. From these values, we note up to nearly as high as 45X reduction (per epoch) in training wait time when the training hardware is switched from IBM Kyoto to IBM Osaka. This (potentially) is more than twice the speedup that is achieved in \cite{resch2021accelerating}.

\subsection{Scalability to larger datasets}
\begin{figure}[t]
    \centering
    \includegraphics[width=\linewidth]{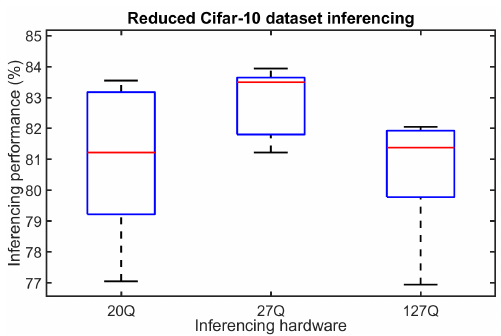}
    \caption{Inferencing performance on various hardware for multi-hardware training on reduced Cifar-10 dataset.}
    \label{fig:cifar_results}
    \vspace{-5mm}
\end{figure}

We also show that our methodology is scalable to larger datasets. In particular, we train a hybrid quantum-classical neural network (with few convolution layers followed by QNN having 8 qubits) using a multi-hardware training setup for the Cifar-10 dataset \cite{cifar10}. We use a reduced version of the dataset consisting of airplane (class 0) and frog (class 6) classes (300 images per class, 70:30 train-test split). Based on the best-scoring configurations from Table \ref{tab:config_analysis}, we select the following configurations: 27Q(II), 20Q(II), 127Q(V), 27Q(I) and 20Q(IV). For all configurations combined, we note mean inferencing accuracy of $80.96\%$, $82.84\%$, and $80.59\%$ for 20, 27, and 127-qubit hardware respectively. Once again, we observe that 27-qubit hardware performs the best, followed by 20-qubit hardware, and finally 127-qubit hardware. We show the boxplots of inferencing performance for various hardware in Fig. \ref{fig:cifar_results}. We also compare these results with another model that is trained only on the best-scoring configuration 27Q(II). From this model, we note mean inferencing accuracy of $82.78\%$, $82.33\%$, and $81.82\%$ respectively for 20, 27, and 127-qubit hardware respectively. From this, we note a mean accuracy reduction of $0.84\%$ across all hardware when we switch from single hardware to multi-hardware training for the Cifar dataset.

\section{Conclusion} \label{conclusion_section}
In this work, we proposed a novel methodology to train QML models on multiple hardware. First, we selected a suitable model followed by a configurational analysis of all configurations. Based on the intuition gained, we chose the top-scoring configurations and proposed a multi-hardware training setup. The results of multi-hardware training show that small datasets such as Iris are resilient to the effect of noise, however, more complex datasets such as Digits images are susceptible to different factors such as coupling constraints and noise characteristics. Finally, we note that the proposed methodology can be scalable to larger datasets such as RGB Cifar-10 images yielding reasonable inferencing performance. 

\section*{Acknowledgements}
We acknowledge the usage of IBM Quantum along with Pennylane for performing all the experiments. All the relevent code has been added to a GitHub Repository\footnote{GitHub repository link: \url{https://github.com/KoustubhPhalak/QuaLITi-QML-Workload-Optimization}}. This work is supported in parts by NSF (CNS-1722557, CNS-2129675,CCF-2210963,CCF-1718474,OIA-2040667, DGE-1723687, DGE-1821766, and DGE-2113839) and Intel’s gift.

\bibliographystyle{IEEEtran}
\bibliography{references.bib}

\end{document}